# Network Analysis of Attitudes towards Immigrants in Asia

Rachael Kei Kawasaki[1*], Yuichi Ikeda[1]

[1]Graduate School of Advanced Integrated Studies in Human Survivability, Kyoto University, 1 Nakaadachi-cho, Yoshida, Sakyo-ku, Kyoto 606-8306
*Correspondence:
kawasaki.kei.52e@st.kyoto-u.ac.jp

**Abstract**:
This study models cross-national attitudes towards immigrants in East and Southeast Asia as a signed and weighted bipartite network of countries and evaluative reactions to a variety of political issues, or determinants. This network is then projected into two one-mode networks, one of countries and one of determinants, and community detection methods are applied. The paper aims to fill two deficiencies in the current research on attitudes towards immigrants: 1) the lack of cross-national studies in Asia, a region where migration is growing, and 2) the tendency of researchers to treat determinants as uncorrelated, despite the interdependent nature of evaluative reactions. The results show that the nine countries in the sample are a cohesive clique, showing greater similarities than differences in the determinants of their attitudes. A blockmodeling approach was employed to identify eight determinants in attitudes towards immigrants, namely views on independence and social dependencies, group identities, absolute or relative moral orientation, attitudes towards democracy, science and technology, prejudice and stigma, and two determinants related to religion. However, the findings of this survey yielded some surprising results when compared with the literature review. First, education was not found to be a significant determinants of attitudes towards immigrants, despite its strong and consistent predictive power in European models. Second, prejudice appears to be mediated in part by religion, especially in religious identification and belief in God. Group identity and prejudice also appear to be related, though only weakly. Finally, anxiety appears in clusters related to social norms, suggesting that fears regarding immigrants relates closely to expectations of others' behavior.

**Keywords**: migration; attitudes; bipartite network; blockmodeling

## 1 Introduction

The past decade has seen migration become a more controversial issue in myriad countries, with stalwart supporters and detractors. Not only does the level of immigration spark impassioned debate, but what happens after immigrants arrive and settle in a country is equally salient. Of particular interest to policymakers and publics is whether immigrants become an accepted part of society, or if they and their children remain at the margins of society.

Attitudes towards immigrants and immigration (ATII)[1] are definitional aspect the integration of immigrants into a society. While integration is a process that includes multiple dimensions, such as labor market inclusion, civil participation, etc., it can also be defined simply as becoming an accepted part of society (Penninx 2007). A large proportion of negative attitudes towards immigrants amongst a country's citizenry by definition shows that immigrants and immigrant groups have not become a fully acceptable part of society.

---

[1] See Ceobanu and Escandell (2010) for a discussion of this term.

Moreover, negative attitudes have practical effects on other aspects of immigrants' integration. Areas where the public has more negative attitudes have been found to have greater housing discrimination against immigrants (Carlsson and Eriksson 2017). Immigrants tend to have lower life satisfaction (Knabe et al. 2013) and greater strains on mental and physical health (Agudelo-Suárez et al. 2009). Investigating why migration is acceptable to some and anathema to others is, thus, of critical importance in ensuring the well-being of all residents, creating a more cohesive society, and guiding policymakers and NGOs in developing more effective and amenable immigration and integration policies.

Given the increasing salience of how attitudes towards immigrants are formed, the current research on the determinants of attitudes towards immigrants has begun to reveal its shortcomings. First, research, especially cross-national research, has been largely limited to European countries and Settler countries, as Table 1 shows. Settler countries here refer to Australia, Canada, New Zealand, and the United States, all former colonies of the United Kingdom which experienced large scale immigration in the 18$^{th}$ and 19$^{th}$ century.

Table 1. Cross-national ATII studies using survey data: locations

| Location | Max. number of countries | Max. number of non-European, non-Settler countries | Data set | Study |
|---|---|---|---|---|
| Europe | 17 | 0 | European Social Survey (ESS) | Meuleman et al. (2009) |
|  | 21 | 0 | ESS | Gorodzeisky and Semyonov (2009) |
|  | 22 | 0 | ESS | Card et al. (2005) |
|  | 22 | 0 | ESS | Hainmueller and Hiscox (2007) |
|  | 26 | 0 | ESS | Malchow-Møller et al. (2009) |
|  | 12 | 0 | Eurobarometer | Gang et al. (2002) |
|  | 15 | 0 | Eurobarometer | Lahav (2004b) |
|  | 17 | 0 | Eurobarometer | Coenders et al. (2005) |
|  | 15 | 0 | Eurobarometer | Kessler and Freeman (2005) |
|  | 12 | 0 | Eurobarometer | Semyonov et al. (2006) |
| North America, South America | 17 | 17 | Latin Barometer | Lawrence (2011) |
|  | 10 | 10 | Latin American Public Opinion Project (LAPOP) | Meseguer and Kemmerling (2016) |
| Africa, Asia, Europe, North America, South America | 31 | 8 | International Social Survey Programme (ISSP) | Facchini and Mayda (2008) |

| Africa, Asia, Europe, North America, Oceania, South America | 53 | 40 | World Values Survey (WVS) | Cooray et al. (2018) |
| --- | --- | --- | --- | --- |
| | 66 | 22 | ISSP; WVS | Mayda (2006) |

There are a several reasons for this overemphasis of European and Settler countries, including an unequal distribution of research resources and capacity in the Global North (Castles 2010) and greater availability and depth of data in European and Settler countries (UN DESA 2017). However, as the destination and origins of migrants change, this lack of understanding of how different countries react to the inclusion of immigrants into a society leads to a poorer ability to make and communicate effective immigration and integration policy. Asia recently overtook Europe as the region hosting the greatest number of migrants in the world and experienced the largest growth in migrant stock in the period between 2000 and 2017 (UN DESA 2017). The current research has thus far failed to reflect these changing conditions.

The overemphasis of Western countries has led to methodological problems and assumptions that compromise the generalizability of research findings and which limit its applicability to Asian countries. These oversights include assumptions of liberal democracies, higher development levels, European and Settler country cultural norms related to ingroups and outgroups, and so on. The effect of this overemphasis is that models and findings are limited to only one region and lose accuracy as the survey expands to other regions (Mayda 2006). Variables that are not relevant to European and Settler countries are overlooked. The scope of the previous research limits the amount of research into structural determinants of ATII, e.g. development level, government corruption, inequality, etc. Without the inclusion of a more diverse group of countries, identifying which determinants are truly generalizable and which are country-specific as well as how macro-level factors affect ATII is difficult.

Finally, the previous research on ATII suffers from a lack of understanding of how variables interact with one another. Often variables are often assumed to be uncorrelated; however research on attitude formation suggests that attitudes are dependent in part on other attitudes (Fazio 1990). Thus, how ATII form and its determinants are a natural area for the application of network science techniques.

This study will identify and analyze the determinants of attitudes towards immigrants in nine East and Southeast Asian countries in a network structure, showing not only which determinants are salient in this critical and understudied region of migration but also the relationships between these determinants. Given that little of the previous cross-national research has been conducted in this region, this study aims to recognize new and understudied trends and variables, with the hope of inspiring greater research into these areas. Using data from the sixth wave of the World Values Survey, countries and variables will first be arranged in a bipartite network, before being projected into one-mode networks. Community detection within the country network will allow for a greater understanding of how countries are similar or unique in their determinants, while blockmodeling of the variable network will allow for the identification of determinants and the relationships amongst them.

This paper is organized as follows: first, the related work on attitudes towards immigrants will be reviewed. Next, the data and methodology used in the study will be described. The results of this methodology and its implications will be discussed, before concluding.

## 2   Related Work

As mentioned above, attitudes form based on a number of considerations including other attitudes. Attitudes towards one object rely on relevant evaluations of other objects, the accessibility of these evaluations, the respondent's affect, mood, and emotion as well as new information, the source of this information, and the respondent's relationship with the source (Crano and Prislin 2006). People can also hold contradicting evaluations of the same object, either through implicit and explicit evaluations that are more or less accessible depending on the situation (Wilson et al. 2000) or through invalidating previous evaluations (Petty et al. 2006). The state of holding conflicting evaluations towards the same object is often referred to as "cognitive dissonance" (Festinger 1957). Which evaluation is expressed depends on the relative strength of each evaluation as well as the motivation of the individual, external conditions, and so on (Wilson et al. 2000). As people try to avoid the discomfort of cognitive dissonance, they may seek out and retain information that supports their view to the detriment of opposing facts (Frey 1989; Brannon et al. 2007; Hart et al. 2009). Thus, attitudes depend on other attitudes, the strength of these attitudes and evaluations, the social network of the respondent, mood, and so on. While some of these factors are outside the scope of the widely available survey data, a respondents' evaluation of other political and social aspects are included in the WVS.

Because of the dependencies between evaluations in creating attitudes, network science has been employed to model the formation of attitudes towards a variety of public opinion issues. One such model is the Causal Attitude Network model (CAN), an empirical application of interrelated evaluative reactions towards an object (Dalege et al. 2016). The network is created by regressing each node, in this case an evaluative reaction towards the object, against another. The parameters of this logistic regression than become the edge weights of the ties between the nodes. This model has been applied to post-national citizenship identities (whether one identifies more strongly as a citizen of the world or of a transnational, supranational identity than as a citizen of their country) using data from 27 European OECD member countries (Schlicht-Schmälzle et al. 2018). While differences between younger and older respondents, rural and urban respondents, etc., are found through a series of Network Comparison Tests, the model does not show differences and similarities amongst countries.

Methodological problems notwithstanding, a wealth of research has been conducted in the determinants of attitudes towards immigrants, and cross-national studies have identified a variety of consistent determinants. Table 2 summarizes and categorizes these factors. Determinants are split into individual-level, both attitudinal and non-attitudinal, and group-level determinants.

Table 2. Summary of determinants of ATII

| Group | | Factors |
|---|---|---|
| **Individual** | Non-attitudinal | - Education<br>- Age<br>- Sex<br>- Rural/urban residence<br>- Economic self-interest |
| | Attitudinal | - Value of cultural homogeneity<br>- Social Dominance Orientation<br>- Satisfaction with democracy in their country<br>- Political orientation<br>- Issue salience<br>- Ethnocentrism/prejudice |

|  |  | - Religion |
|---|---|---|
| **Group** |  | - Group contact<br>- National identity |
| **Macro** |  | - National economic conditions<br>- Local immigrant population<br>- Policy |

Macro-level factors seek to describe the economic and social environment in which individuals and groups make their decisions. These factors are of particular interest to policymakers as they fall more within the purview of policy than the personal attitudes of individuals. Often, macro-level factors are found to have a mediating effect. Individuals' perception of the overall condition of the national economy, rather than their personal economic circumstances, have been found to be more predictive of attitudes towards immigrants (Coenders et al. 2008; Lahav 2004; Semyonov et al. 2008), with varying effects based on the country's development level (Mayda 2006; Kunovich 2004). However, within the literature, there is little consensus on how the size of the local immigrant population affects attitudes some studies have found that larger local immigrant population may encourage more positive attitudes towards immigrants (Fetzer 2000; Ellison et al. 2011), others a negative relationship (Semyonov et al. 2006; Semyonov et al. 2008; Jackson et al. 2001), and some finding no relationship at all (Hello et al. 2002; Hjerm 2007; Sides and Citrin 2007).

The political system of a country also appears to play a mediating role on attitudes. In long-established liberal-democracies, education has a more liberalizing effect on people's attitudes towards immigrants than in countries with a disrupted or shorter history of democracy (Coenders and Scheepers 2003; Hello et al. 2002). Finally, policy can have an effect on attitudes by reframing the debate around immigration. Public opinion can become more relaxed or more stringent in accordance with more inclusive or more restrictive policies (Calavita 1996; Chavez 2008; Flores 2014; Campbell 2012).

While this study cannot conclude on the influence of macro-level factors on attitudes, it does propose some hypotheses on the effect of migration policy and economic and political structure on ATII. The following section will highlight factors which are relevant to the study.

## 2.1 Individual-level: non-attitudinal

One of the most commonly cited determinants of ATII in the popular conversation are the respondent's personal economic situation, whether one stands to win or lose economically from the inclusion of immigrants in the labor market. The empirical evidence, however, show little support for this theory. Scholars posit that economic self-interest shapes ATII through two main mechanisms: by affecting a person's relative wages, as described above, or by affecting a person's fiscal burden. The effect of wages and income have been found to have a relationship in some studies (Mayda 2006; Coenders et al. 2008; Jackson et al. 2001; Kehrber 2007) and been found to be insignificant in others (Semyonov et al. 2006; Semyonov et al. 2008; Wilkes et al. 2007). Fiscal burden theories posit that an individual's ATII is determined by whether they believe that immigrants will be a net burden on the welfare state and increase the respondent's taxes as a result (Campbell et al. 2006; Dustman and Preston 2007). Again, studies disagree on its effects, with some finding that respondents with higher incomes were found to be less supportive of immigration (Facchini and Mayda 2008) and others finding the opposite (Kunovich 2004)

However, an individual's economic standing often shows less predictive power once stereotypical thinking about immigrant groups is accounted for (Burns and Gimbel 2000).

Hainmueller and Hopkins go so far as to categorize the economic self-interest as "a zombie theory" (2014, pp. 241), as it continues to live on without strong empirical support. Rather, the theoretical simplicity and precision of self-interest theory "may help explain the staying power of that approach in the face of both observational and experimental results that favor symbolic threats" (Hainmueller and Hopkins 2014, pp. 242).

Education has consistently been found to be a significant and powerful predictor of attitudes towards immigrants and immigration in many cross-national studies (Lancee and Sarrasin 2015; Freeman et al. 2013). As level of education increases, especially to or past the tertiary level, support for immigration increases and anti-immigrant sentiment decreases. However, the mechanism through which education exerts influence on ATII is still debated. From a political economy perspective, workers with higher education are higher skilled and, therefore, not in competition with immigrants, who tend to be low-skilled (Scheve and Slaughter 2001; Mayda 2006; O'Rourke and Sinnott 2006). However, many have noted that education correlates with a range of other variables which affect ATII (Coenders and Scheeper 2003; Hainmueller and Hopkins 2014), such as reduced authoritarianism (Hello et al. 2006) and an emphasis on cultural diversity (Davidov and Meuleman, 2012). These cultural value correlates may stem from selection biases (Lancee and Sarrasin 2015), or university education may encourage greater reflexivity and critical thinking, dismantling more dichotomous or simplified beliefs about differences between groups (Gang et al. 2002; Chandler and Tsai 2001). Equally, universities may foster more contact with different cultures and cause students and graduates to have more diverse social circles and reduced prejudice (Case et al. 1989), a theory known as group contact theory (Allport 1954; Fetzer 2000; Schlueter and Wagner 2008). In few cross-national studies which takes place outside of these European and Settler countries, education's remains significant but to a lesser extent. In Meseguer and Kemmerling's (2016) study in 10 Latin American countries of varying development levels, also found a limited effect of education on attitudes. The effect of education on attitudes lost significance when including other variables such as employment status and occupation. The authors conclude that education is a poor indicator of labor market competition and has limited predictive power in Latin American countries.

## 2.2 Individual-level: attitudinal

As mentioned, attitudinal factors, especially factors which measure prejudice against minorities generally or certain minority groups, ethnocentrism, and political orientation, have been found to have significant effects on ATII. A study of 20 European countries found that the degree to which a person values cultural homogeneity is a much more accurate predictor of their attitudes towards immigrants than their economic standing (Sides and Citrin 2007). Worries about diminished national unity and uniqueness were also found to outweigh concerns about a decay of national sovereignty or media influence on attitudes (Ivarsflaten 2005). In a study conducted with American undergraduate students, respondents who rate higher on a social dominance scale, the degree of preference for hierarchy and inequality between social groups and, as such, is predictive of a range of political and social attitudes (Pratto et al. 1994), were more angered by encounters with Spanish language and had more negative attitudes towards immigrants (Newman et al. 2013). Political orientation has also been found to be significant in determining people's ATII. In general, people who identify with the political right are more likely to support restrictive immigration policies (Hainmueller and Hiscox 2007; McLaren 2001; McLaren 2003; Semyonov et al. 2006; Semyonov et al. 2008) and favor excluding immigrants from equal access to institutions as native-born citizens, such as healthcare or welfare (Gorodzeisky and Semyonov 2009). Furthermore, in some studies, the effect of political orientation remains even after controlling for other factors (Kunovich 2009; Semyonov et al. 2006; Knoll et al. 2011).

A study of 15 European Union member states finds that one's satisfaction with the democratic process in one's country was also found to have an inverse relationship with ATII (Weldon 2006). Those who were more dissatisfied with the democracy of their country were more likely to support exclusionary policies towards immigrants, including limiting their rights to free speech, religious freedoms, and voting rights (Weldon 2006). More generally, higher institutional trust is correlated with more inclusive attitudes towards immigrants, as a more recent cross-national study of European countries shows (Halapuu et al. 2013). A similar, single-country study found the same trend in England (Andreescu 2011).

Similarly, one's affect and general anxiety has been found to be significant in determining ATII. In one experiment on information processing and immigration, respondents were made to feel anxious about immigration and were then asked to search for information about immigration on a website containing an equal number of threatening and nonthreatening information. Respondents who had been primed to feel anxious sought out more threatening information and were better able to recall threatening information (Gadarian and Albertson 2014). Respondents were asked what specific concerns caused them to feel anxious about immigration. These responses could be classified into the following four categories: economic concerns, concerns about exploiting the social welfare system, cultural worries, and security concerns (Gadarian and Albertson 2014).

Prejudice against other ethnicities has been found overwhelmingly to be linked towards anti-immigrant prejudice, though a central question remains as to whether prejudice against immigrants is origin-blind (Citrin et al. 1997; Sniderman et al. 2004; Kinder and Kam 2009) or if people differentiate their attitudes towards immigrants based on country of origin, ethnicity, and cultural distance. Citrin et al. finds that personal economic standing had little predictive power in a person's attitudes towards immigrants, but that negative attitudes towards Latinos and Asian Americans and pessimism about the national were significant (Citrin et al. 1997). Sniderman et al. find that Italian respondents with negative attitudes towards African immigrants tend to be equally against Eastern European immigrants, supporting the theory that prejudice towards immigrants are origin-blind (Sniderman et al. 2000). Looking at the United Kingdom, Ford finds a consistent preference for white and culturally similar immigrant groups using the British Social Attitudes survey from 1983 to 1996 (Ford 2011). Hainmueller and Hangartner exploit the natural experiments of Swiss local elections which, prior to 2003, could include referendums to determine which applicants should be granted Swiss citizenship (Hainmueller and Hangartner 2013). Thus, the research provides no consistent answer as to whether individual prejudice is dependent on the immigrant groups' country or origin or not.

Finally, religion has been found in various studies to be significant in determining attitudes but the research is inconclusive, as religion has been found to be have both positive (Scheepers et al. 2002) and negative impact on anti-immigrant sentiment (Mayda 2006). Studies often differ in how they define and measure religion, with some studies looking at which specific sect people ascribe to, e.g. Catholic, Protestant, Muslim, and others measuring church attendance or a subjective measure of how important religion is to the respondent (Scheepers et al. 2002). Thus, results are not always comparable. In a cross-national study, Scheepers et al. find that religion exerts a similar effect across countries, but how the individual expresses or understands their religion greatly influences the result (Scheepers et al. 2002). Those who subscribed to a formal religion were found to have greater prejudice than those who were non-religious. Christians, both Catholics and Protestants, showed the most anti-immigrant sentiment. Moreover, those who expressed a greater level of religious particularism or valued doctrinal adherence were also more likely to express prejudice against ethnic minorities. However, respondents who answered that religiosity was more important in their lives and who felt more strongly that they had a spiritual life (as compared to an intellectual or emotion life) were found to have lower levels of prejudice (Scheepers et al., 2002).

## 2.3 Group-level

Group-level factors are factors which relate to an individual's attachment to a group and where the primary concern is the group's interests rather than the individual's self-interest. Studies which advance these theories generally rely on two theoretical foundations: the concept of intergroup conflict and the concept of social identity theory. Intergroup conflict theory states that prejudice between groups arises due to real or perceived conflicts over a limited amount of resources (Esses et al. 1998). These resources can be divided into two categories, material resources and symbolic resources. Symbolic resources in this case can refer to esteem, social value, and other social rewards (Smith et al. 2015). This theory is often marshalled in support for economic theories as the primary motivation for negative ATII. However, as symbolic resources can be an equal area of conflict, intergroup conflict can be applied to cultural and attitudinal factors of ATII on the group level. Finally, intergroup conflict also stresses that the perception of threat to a group resource can be as powerful as the actual conflict.

In contrast to intergroup conflict theory, social identity theory centers the primacy of identities in shaping opinions towards others and other groups. It states that individuals categorize other people into either an outgroup (for those who are determined to be different from themselves) and an ingroup (for themselves and those who are similar to them). Limiting the opportunities of those in the outgroup becomes one strategy to increase the distinctiveness and cohesion of the ingroup (Esses et al. 1998). Through the exclusion of members of the outgroup, the ingroup can fend off perceived threats to their symbolic resources, in this case the positive identity gained from the group. Therefore, those who identify more heavily with a group are likely to have more hostile feelings towards outgroups. Social identity theory posits that people who more strongly identify with the nation and who perceive threat more strongly will have a more negative reaction towards immigrants. National identity, meaning the degree to which a person is attached to their country, has been found to have a positive relationship with more exclusionary attitudes towards immigrants and immigration (Kunovich 2009; Coenders and Scheepers 2004; Mummendey et al. 2001; Gijberts and Hagendoorn, 2017).

## 3 Data

This study uses data from the sixth wave of the World Values Survey, conducted between 2010 to 2014 (Inglehart et al. 2014). The World Value Survey seeks to measure and track "the beliefs, values and motivations of people throughout the world" (WVS Database 2020) and asks respondents over 200 questions related to topics such as religion, governance, social relations, morality, and so on. The survey employs a common questionnaire that is translated by a national team composed of social science researchers from universities and/or research institutes. The national team then conducts a nationally representative survey of residents 18 and older, regardless of their nationality, to reach a minimum sample size of at least 1200 people (Inglehart et al. 2014). The data from the following nine countries and territories, four in Southeast Asia and five in East Asia, was used in this study: China, Hong Kong, Japan, Malaysia, the Philippines, Singapore, South Korea, Taiwan, and Thailand. Immigrants were eliminated as the study aims to understand the attitudes and reactions of autochthonous populations towards immigrants. China, Hong Kong, and Japan did not record if any respondent was an immigrant. Finally, those who did not respond to the dependent variable were also eliminated from the sample. In total, the combined number of respondents for this study was 12,119.

The dependent variable asks respondents about prioritizing autochthonous people in the hiring process. The question was recoded so that a response [1] would indicate a more exclusionary attitude towards immigrants and [0] would indicate a more lenient or inclusive attitude. The survey questions is as follows:

V46. When jobs are scarce, employers should give priority to people of this country over immigrants.

[1] Agree
[0] Neither *originally coded [2]
[0] Disagree *originally coded [3]

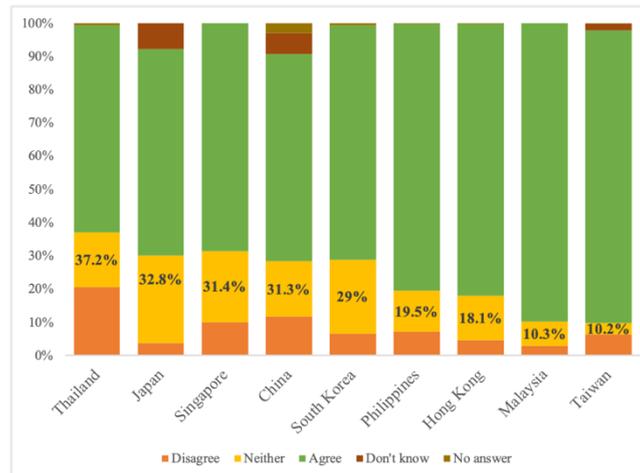

Figure 1. Responses to independent variable for nine countries
Percentages represent percentage of people who disagreed or neither agreed or disagreed with the independent variable, study response coded (0). Data source: Inglehart et al. 2014.

## 4  Methodology

Table 3 shows a summary of the methodology used in this study. The methodology is divided into two distinct stages: pre-selection and network analysis. The pre-selection applies a series of statistical tests in order to determine which variables are the most significant in determining attitudes towards immigrants in the nine countries. Once these variables are identified, it was possible to create a bipartite network with a class of country nodes and a class of variable nodes, showing the relationship between independent variables and the dependent variable for each country. This bipartite network was then projected, creating two one-mode networks, one country network and one variable network. As the country network consists of only nine nodes, Louvain clustering was used to detect communities within the network. The variable network consisted of a total of 45 variables and contained both negative and positive edges. For this network, blockmodeling was applied to find the determinants, or clusters of highly correlated variables, of attitudes towards immigrants.

Table 3. Stages, methods and their description of the study

| Stage | Methodology | Description |
|---|---|---|
| Pre-selection | 1. Spearman's correlation | |
|  | 2. Chi-squared test of independence | |
|  | 3. Linear regression of continuous variables | |
|  | 4. Multiple logistic regression | |
| Network Analysis | 1. Link weight in bipartite network | i. Logistic regression |

| | 2. One-mode projection: country network | i. Normalization<br>ii. Matrix multiplication<br>iii. Community detection: Louvain method |
|---|---|---|
| | 3. One-mode projection: variable network | i. Normalization<br>ii. Matrix multiplication<br>iii. Community detection: Blockmodeling |

## 4.1 Pre-selection method

Given the wide range of questions included in the World Values Survey, it was necessary to eliminate variables which were not significantly related to the dependent variable before conducting network analysis. Previous studies generally pre-select variables based on the theoretical or empirical literature. However, in order to allow for the inclusion of variables which may have been overlooked by previous studies or are less relevant in European and Settler countries, this study pre-selects variables by conducting a series of statistical tests which seek to identify variables which have a significant relationship with the dependent variable. Variables were divided into three categories: demographic variables, the education variable, and cultural variables. Demographic and education variables were taken from the model proposed by Mayda (2006). Demographic variables included variables for subjective social class, subjective income, sex, age, town size, and political orientation. The education variable measured the highest variable attained. Cultural variables were subjected to a series of statistical tests. A summary of the pre-selection process is provided in Table 3. The final step of the pre-selection process was a multiple logistic regression including demographic, education, and cultural values, with variables deleted according to backwards deletion, until all variables showed significance. Significant variables were included in the network analysis. The pre-selection method was repeated for each country individually, to find the variables most relevant to attitudes towards immigrants in each country.

First, the pre-selection sought to eliminate variables that were highly correlated with one another. For pairs of cultural variables which were highly correlated, the variable which had a less significant relationship with the dependent variable was eliminated. Variables, $X = \{X_i\} (i = 1, \cdots, N)$ and $Y = \{Y_i\} (i = 1, \cdots, N)$, were converted to rank variables $r_X$ and $r_Y$. $N$ is the number of the responses (answers). Variables, $X$ and $Y$, were considered highly correlated if

$$0.75 \leq |cor(r_X, r_Y)| = \left|\frac{cov(r_X, r_Y)}{\sigma_{r_X} \sigma_{r_Y}}\right| \qquad (1)$$

where $cor$ is the correlation of rank variables $r_X$ and $r_Y$, or is equal to the covariance of the rank variables divided by the standard deviation $\sigma_{r_X}$ and $\sigma_{r_Y}$. If a pair of variables were found to be highly correlated, then logistic regression against the dependent variable was used to determine which variable, $X$ or $Y$, had a more statistically significant relationship with the dependent variable $P$.

$$\log\left(\frac{P_i}{1-P_i}\right) = \alpha^{(0)} + \alpha^{(1)} X_i + \varepsilon_i \qquad (2)$$

where $P = \{P_i\}$ $(i = 1, \cdots, N)$ is the probability that the dependent variable $Y$ will equal 1,. Statistical significance with the dependent variable was judged by the Akaike Information Criterion (AIC) of the logistic regression model, as defined by

$$AIC = -2\ln(L) + 2s \tag{3}$$

where $L$ is the value of the likelihood and $k$ is the number of estimated parameters. The variable for which the model's AIC was smaller was retained, while the other variable in the pair was eliminated. As both variables have an equal number of parameters ($s = 1$), the variable with the larger $L$, was selected.

After eliminating highly correlated variables, a chi-squared test of independence was used to eliminate variables that did not show a statistically significant difference between the expected frequencies and observed frequencies of responses to the dependent variable. The chi-squared test was calculated according to the formula

$$\chi^2 = \sum_{i=1}^{M} \frac{(O_i - E_i)^2}{E_i} \tag{4}$$

where $O = \{O_i\}$ ($i = 1, \cdots, M$) is the observed frequency of responses to the dependent variable for each response, and $E = \{E\}$ ($i = 1, \cdots, M$) is the expected frequency of responses to the dependent variable for each response. Here $M$ is the number of possible selection of the responses. Under the assumption of no association, the null hypothesis, the probability of negative attitudes towards immigrants would be similar regardless of the response to the independent variable. Variables for which the p-value of $\chi^2$ was less than or equal to 0.05, given the degrees of freedom, were retained, while variables where the null hypothesis could not be rejected were eliminated.

Following the chi-squared test, continuous variables, defined as variables with more than three possible selection of the responses, were subjected to an additional test. The aim of this test was to eliminate variables for which there was not a linear relationship with the dependent variable, supplementing the chi-squared test. In this case, the frequency of exclusionary attitudes towards immigrants for each response to the dependent variable was regressed against the responses to the dependent variable, $Y$. $Y_i$ was the proportion of exclusionary attitudes for each response $i$, to the dependent variable. For each

$$Y_i = \beta^{(0)} + \beta^{(1)} X_i + \varepsilon_i \tag{5}$$

where $X_i$, and $\beta$ are the parameters of the model. If the p-value of the model was greater than 0.05, the dependent variable was eliminated.

Finally, demographic, education, and cultural variables were regressed against the dependent variable in a multiple logistic regression.

$$\log\left(\frac{P_i}{1-P_i}\right) = \gamma^{(0)} + \sum_{j=1}^{V'} \gamma^{(j)} X_i^{(j)} + \varepsilon_i \tag{6}$$

where $P = \{P_i\}$ ($i = 1, \cdots, N$) is the probability that the dependent variable $Y$ will equal 1, given the dependent variable $X$, and $\gamma$ are the parameters of the model. Variables were eliminated via stepwise backwards deletion according to highest p-value. When all variables in the multiple logistic regression were significant above 0.05, these variables were selected for the network analysis.

### 4.2 Network analysis

Having selected the variables, the weights of the edges between countries and variables were obtained by logistic regression. Variables were rescaled according to min-max normalization, as independent variables had different ranges of responses, from binary questions to questions with up to ten possible responses.

$$X'_i = \frac{X_i - \min(X)}{\max(X) - \min(X)} \qquad (7)$$

For a variable $X$, each element $X_i$ was linearly transformed to $X'_i$, so that responses ranged from 0 to 1.

A bipartite network was then constructed, with a class of nodes for countries and a class of nodes for variables. The weight of the edges between the classes is determined by the regression coefficient of the independent variables on the dependent variable for each country. In short, the bipartite network shows the effect of each variable on the dependent variable for each country c. The edge weights, $\delta_c^{(j)}$ between country node c and variable node j were obtained by the following formula:

$$\log\left(\frac{P_i}{1-P_i}\right) = \delta^{(0)} + \delta_c^{(j)} X_i^{(j)} + \varepsilon_i \qquad (8)$$

where $P = \{P_i\}$ $(i = 1, \cdots, N)$ is the probability that the dependent variable $Y$ will equal 1, and $\delta$ are the parameters of the model. If the p-value of the model is greater than 0.05, then the edge weight is equal to 0.

Using the parameters of the simple logistic regression, a bipartite weighted adjacency matrix was created, $A = \{a_{ij}\} = \{\delta_i^{(j)}\}$ $(i = 1, \cdots, C; j = 1, \cdots, V)$. Here $V$ is the number of selected variables. The edge weights $a_{ij}$ were created by the parameters of the simple logistic regression. This weighted adjacency matrix is shown in the appendices.

In order to understand the relationships between variables within the same class, i.e. the relationships between countries (variables), it is necessary to make two one-mode projections of the network, first of the country class and then of the variable class. These one-mode projections multiply country (variable) vectors by other country (variable) vectors to produce a scalar value representing the similarity between the two vectors.

First, the matrix was rescaled so that after matrix multiplication, the dot product between two countries or two variables would be on a scale from -1 to 1. Matrix A is rewritten using row-wise country vector $c^{(i)}$ $(i = 1, \cdots, C)$:

$$A = \begin{bmatrix} c^{(1)} \\ c^{(2)} \\ \vdots \\ c^{(C)} \end{bmatrix} = \{c^{(i)}{}_j\} \qquad (9)$$

Each country vector $c^{(i)}$ was normalized according to the following formula:

$$\hat{c}^{(i)} = \frac{c^{(i)}}{\sqrt{\sum_{j=1}^{V} c^{(i)}{}_j{}^2}} \qquad (10)$$

Having normalized the country vectors, the normalized matrix

$$\widehat{A} = \begin{bmatrix} \hat{c}^{(1)} \\ \hat{c}^{(2)} \\ \vdots \\ \hat{c}^{(C)} \end{bmatrix} = \{\hat{c}^{(i)}{}_j\} \tag{11}$$

was multiplied by its transposed normalized matrix $\widehat{A}^T$, in order to find the degree of similarity between countries $\Phi = \{\phi_{ij}\}$ $(i = 1, \cdots, C; j = 1, \cdots, C)$.

$$\Phi = \widehat{A} \cdot \widehat{A}^T \tag{12}$$
$$\phi_{ij} = \sum_{k=1}^{V} \hat{c}^{(i)}{}_k \hat{c}^{(j)}{}_k \tag{13}$$

The result of this operation is a weighted adjacency matrix of the one-mode projection of countries $\{\phi_{ij}\}$, shown in the appendix

The same process was then repeated to find which independent variables showed greater similarity. Matrix A is rewritten using column-wise variable vector $v^{(j)}$ $(j = 1, \cdots, V)$:

$$A = [v^{(1)} \quad v^{(2)} \quad \ldots \quad v^{(V)}] = \{v_i^{(j)}\} \tag{14}$$

Each variable vector $v^{(j)}$ was normalized according to the following formula:

$$\hat{v}^{(j)} = \frac{v^{(j)}}{\sqrt{\sum_{j=1}^{C} v_i^{(j)^2}}} \tag{15}$$

Having normalized the variable vectors, the normalized matrix

$$\widehat{B} = [\hat{v}^{(1)} \quad \hat{v}^{(2)} \quad \ldots \quad \hat{v}^{(V)}] = \{\hat{v}_i^{(j)}\} \tag{16}$$

multiplies its transposed normalized matrix $\widehat{B}^T$, in order to find the degree of similarity between countries $\Psi = \{\psi_{ij}\}$ $(i = 1, \cdots, V; j = 1, \cdots, V)$.

$$\Psi = \widehat{B}^T \cdot \widehat{B} \tag{17}$$
$$\psi_{ij} = \sum_{k=1}^{C} \hat{v}_k^{(i)} \hat{v}_k^{(j)} \tag{18}$$

This produced the weighted adjacency matrix of the one-mode projection of variables $\{\psi_{ij}\}$, shown in the appendices

Following the one-mode projection of the countries, a Louvain clustering was applied in order to detect communities within the network (Blondel et al., 2008). The Louvain method optimizes modularity, which is defined as

$$Q^{(C)} = \frac{1}{2m^{(C)}} \sum \left[ \phi_{ij} - \frac{k^{(C)}{}_i k^{(C)}{}_j}{2m^{(C)}} \right] \delta(c_i c_j) \tag{19}$$
$$Q^{(V)} = \frac{1}{2m^{(V)}} \sum \left[ \psi_{ij} - \frac{k^{(V)}{}_i k^{(V)}{}_j}{2m^{(V)}} \right] \delta(c_i c_j) \tag{20}$$

Here $\phi_{ij}$ represents the edge weight between nodes $i$ and $j$, $k^{(C)}{}_i$ is the sum of weights attached to node $i$, $k^{(C)}{}_j$ is the sum of weights attached to node $j$, $m^{(C)}$ is the sum of edge weights in the country network. $\psi_{ij}$ represents the edge weight between nodes $i$ and $j$, $k^{(V)}{}_i$ is the sum of weights attached to node $i$, $k^{(V)}{}_j$ is the sum of weights attached to node $j$, $m^{(V)}$ is the sum of edge weights in the variable network. The Kronecker's delta is defined as follows

$$\delta = \begin{cases} 1 & (c_i = c_j) \\ 0 & (c_i \neq c_j) \end{cases} \qquad (21)$$

where $c_i$ is the community of node $i$ and $c_j$ is the community of node $j$.

Because of the presence of negative ties between variable nodes in the one-mode projection, Louvain clustering could not be used. For this reason, blockmodeling was employed instead. Blockmodeling is a technique which permutes the order of nodes in an adjacency matrix to find clusters. As the network is signed, Doreian and Mrvar's relaxed structural balance blockmodel, a generalization of structural balance blockmodels, was used (2009). The relaxed structural balance blockmodel also seeks to optimize a criterion function. In the case of this study, the following criterion function was used:

$$P(C_k) = \rho \mathcal{N} + (1 - \rho)\mathcal{P} \qquad (22)$$

$P(C_k)$ equals the sum of inconsistencies which violate structural balance in the network, given $k$ number of clusters, $C$. $\mathcal{P}$ represents the total number of positive ties in a negative block, $\mathcal{N}$ the total number of negative ties in positive blocks, and $0 \leq \rho \leq 1$. This criterion function resembles Harary et al.'s (1965) line index of imbalance; however, $\rho$ allows for positive ties in negative blocks and negative ties in positive blocks to be weighted differently. When $\rho = 0.25$, negative and positive inconsistencies are equally weighted. In the case of this study, $\rho = 0.75$, as the aim was to prioritize consistency within blocks over consistency between blocks.

As the criterion function decreases monotonically (Doreian and Mrvar, 2009, pg. 5), it is necessary to dictate the number of clusters. In order to determine the number of clusters, the eigenvectors of the correlation matrix of the variables were taken. Using the eigenvalues, it was possible to determine the number of eigenvectors, $n$, were necessary to explain a majority of the variance in the model. Given that for each eigenvector a variable could have either a positive or negative sign, $k$ number of clusters would be sufficient to explain a majority of the variance in the model.

$$k = 2^n \qquad (23)$$

Finally, to determine the meaning of each block, the node strength of the variable within the block was taken. The node strength is the sum of the weights of ties from a variable to other variables within its cluster. Through this method, the most central variable within the cluster was found. A keyword was then chosen by examining the survey question of the most central variable and the other variables in the block and referring to the classifications provided by the World Values Survey. In several blocks, the node with the highest node strength was part of a series of survey questions which also appear in the block. These series are often used to measure a general tendency, like prejudice or group identification. In this case, the keyword was taken from the series of questions.

# 5 Results
## 5.1 Countries

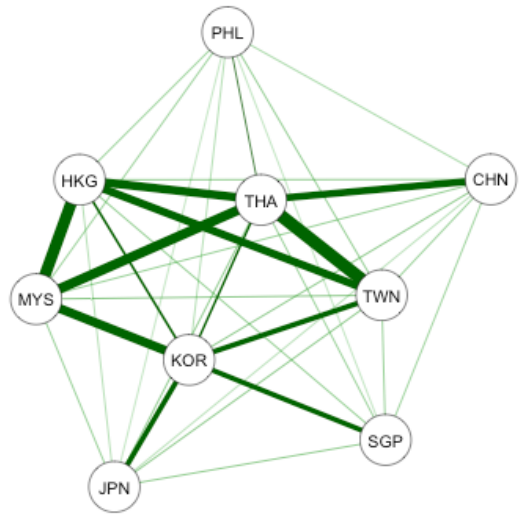

Figure 2. Country network.
Green links represent positive ties. Edge weight represented by width of edge.

Figure 2 shows the network of countries that is the result of the one-mode projection of the bipartite graph. Green edges represent positive ties between countries, and the weight of the edges are represented by the width of the edge. The graph is a clique, meaning that all countries have ties with all other countries. All edges are positive, meaning that countries are similar in their determinants of attitudes. A Louvain clustering algorithm was applied to detect any communities within the network. Dividing the network in two clusters resulted in a negative, though negligible, modularity score (-2.046974e-16), meaning that keeping the network as a single cluster is the optimal partition of the network.

     Some of the edges between countries have larger weights, or show a greater degree of similarity between countries. Summing the weights of ties from a country (node strength) can help illustrate which countries show a large degree of similarity to the rest of the countries in the sample and which countries are fairly unique.

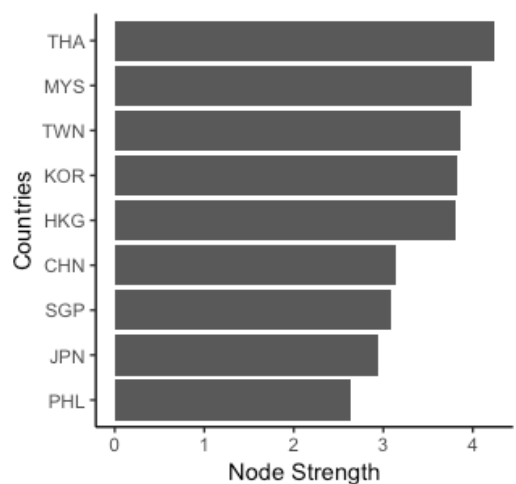

Figure 3. Node strength of country network

As shown by Figure 3, countries Thailand, Malaysia, Taiwan, and Korea have the largest node strength. The Philippines had the lowest node strength, meaning it is the most unique country with regards to its determinants of attitudes.

## 5.2 Variables

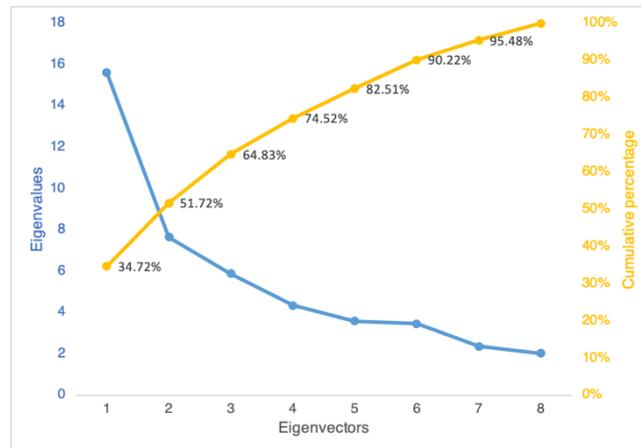

Figure 4. Eigenvalues and Cumulative Variance per Eigenvector in Variable Network.

Figure 4 shows the cumulative percentage of variance for each eigenvector of the correlation matrix of the variable network. Given the cumulative percentage of variance, three eigenvectors are sufficient to explain 64.83% of the variance in the model. As each variable could be either negatively or positively signed in each vector, this suggests that eight clusters ($2^3$), would be sufficient to explain 64.83% of the variance of the model.

Having specified that eight clusters would be sufficient, the following blockmodel (Figure 5) was created.

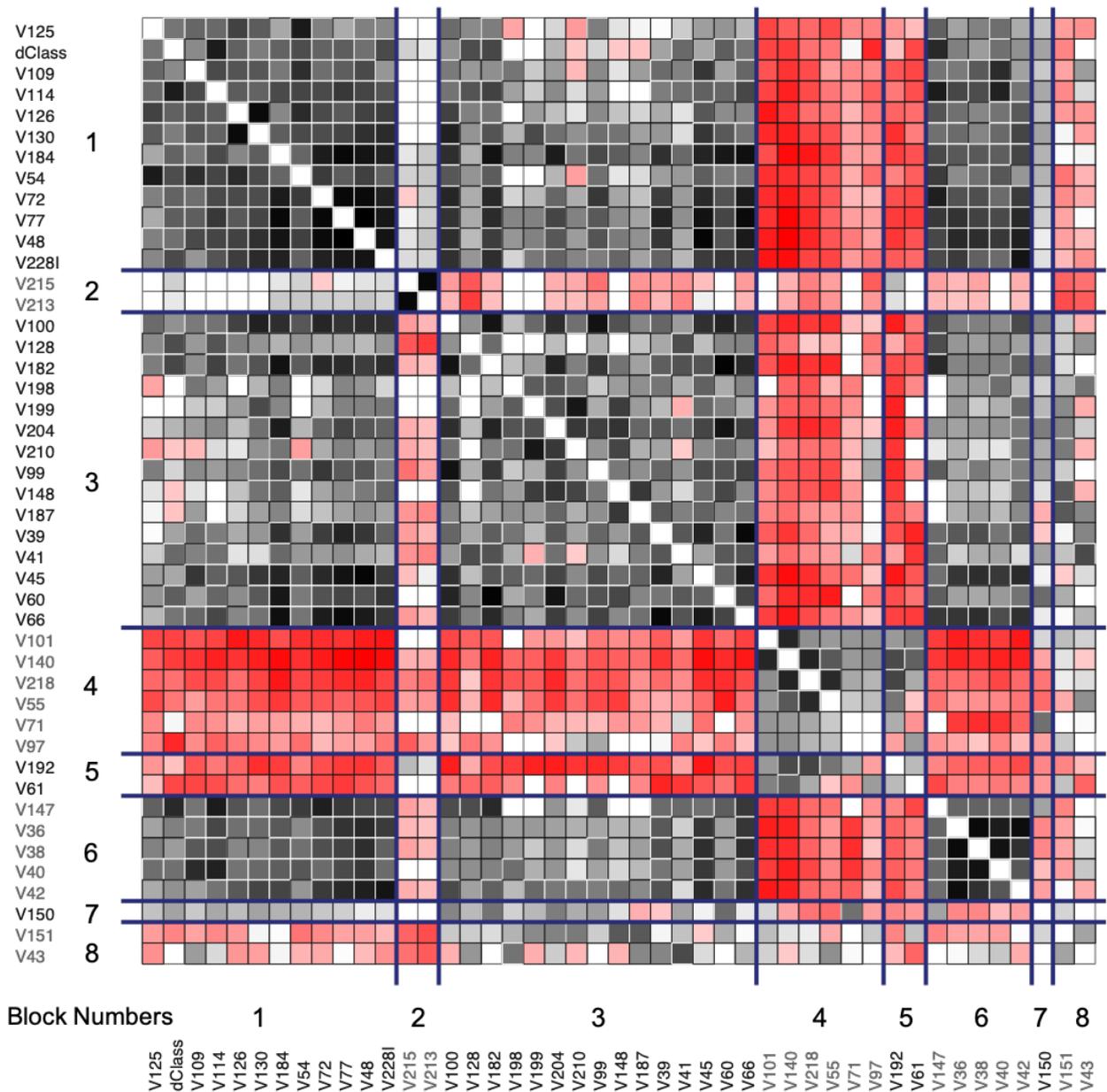

Figure 5. Blockmodel of variable network.
Black squares represent positive ties; red squares represent negative ties. Darker shades represent greater similarity; lighter shades represent less similarity. Created in Pajek.

The blockmodel created eight blocks of variables, henceforth called determinants. Between variables clustered in the same blocks, only two negative ties were found. Both of these negative ties were found in Block 3, one between V41 and V210 and the other between V41 and V199. The total inconsistency score for the partition, or *P(C)*, was 4.930.

When the variables in the variable network are clustered by block, the network shown in Figure 6 is created.

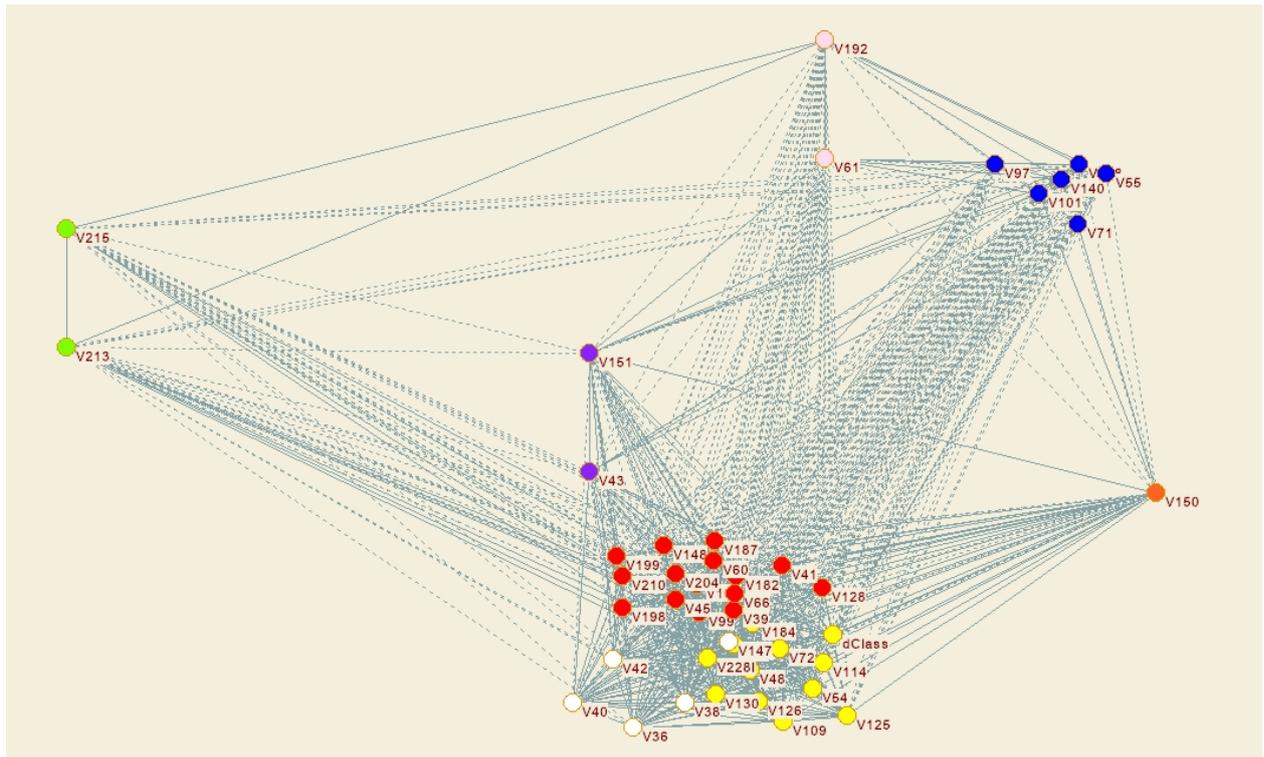

Figure 6. Variable network with clusters from blockmodel.
Colors represent clusters dictated by blockmodel. Positive ties represented by solid lines; negative ties represented by dashed lines.

Finally, the blockmodel shown in Figure 5 can be reduced to an image matrix, showing the overall signs of the blocks. This image matrix is shown in Table 4. "P" represents positive blocks, in which the majority of blocks are positive, and "N" represents negative blocks. Positive blocks off the diagonal are considered an error when structural balance is not relaxed and are filled with grey.

Table 4. Image matrix of blockmodel

|   | 1 | 2 | 3 | 4 | 5 | 6 | 7 | 8 |
|---|---|---|---|---|---|---|---|---|
| 1 | P | P | P | N | N | P | P | N |
| 2 | P | P | N | N | P | N | P | N |
| 3 | P | N | P | N | N | P | P | P |
| 4 | N | N | N | P | P | N | N | P |
| 5 | N | P | N | P | P | N | N | N |
| 6 | P | N | P | N | N | P | N | N |
| 7 | P | P | P | N | N | N | P | P |
| 8 | N | N | P | P | N | N | P | P |

Having described the overall structure, each block was then examined in more detail. The following tables provide the survey question to which each variable refers, ordered from greatest to smallest node strength. The keyword to define each determinant is also described.

The scalar value of similarity between variables, as obtained by the one-mode projection, is sometimes discussed. These values can be found in Appendix G. Furthermore, the relationship between the dependent variable and variables in each country can be found in Appendix E.

Block 1 represents variables related to independence and social dependencies. The cluster also includes questions related to normative behavior (V77), confidence and trust in various institutions (V228I, V126, V114, V109, V125), and a variable related to class. In addition, V72 and V184 relate to security and anxiety amongst respondents.

Table 5. Survey questions: Block 1 – Independence

| Variable | Survey Question | Node Strength |
|---|---|---|
| V48 | Having a job is the best way for a woman to be an independent person | 8.224 |
| V72 | Living in secure surroundings is important to this person; to avoid anything that might be dangerous. | 7.897 |
| V77 | It is important to this person to always behave properly; to avoid doing anything people would say is wrong. | 7.881 |
| V54 | [Do you agree that] being a housewife is just as fulfilling as working for pay? | 7.719 |
| V228I | Voters are offered a genuine choice in elections | 7.694 |
| V126 | [H]ow much confidence [do] you have in [the United Nations] | 7.446 |
| V130 | [Is having a democratic political system] a very good way of governing a country | 7.443 |
| V114 | [H]ow much confidence [do] you have in [the courts] | 7.441 |
| V184 | To what degree are you worried about [a terrorist attack] | 7.213 |
| dClass | Would you describe yourself as belonging to the [upper class, upper middle class, lower middle class, working class, lower class] | 6.543 |
| V109 | [H]ow much confidence do you have in [the armed forces] | 6.464 |
| V125 | [H]ow much confidence you have in [regional organizations] | 5.902 |

As can be seen from Figure 7, Block 1 is a clique, meaning that all the nodes are connected by edges to all other nodes. All ties are positive, showing that the variables are correlated with other variables in the block.

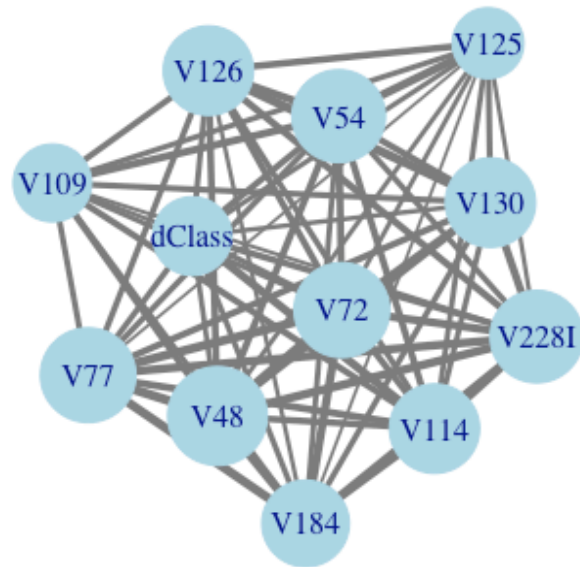

Figure 7. Network of variables in Block 1
Positive ties represented by grey links; negative ties by red ties.

Block 2 contains two variables from a series regarding identification with a group. Respondents were asked whether they see themselves as part of their local community (V213) and part of larger, political organizations, namely their regional organizations (V215). As Block 2 only contains two variables, the node strength of the variables are equal. Similarly, as Block 2 only contains one edge, no network graph was drawn, as a graph would provide no additional information.

Table 6. Survey questions: Block 2 – Group Identity

| Variable | Survey Question | Node Strength |
|---|---|---|
| **V213** | I see myself as part of my local community | 0.925 |
| **V215** | I see myself as part of the [regional organization] | 0.925 |

Block 3 contains a series of questions related to what degree certain infractions can be justified depending on the circumstances or if these infractions are never justifiable. This series asks the respondent to make moral judgements on certain situations, allowing the respondents to express either a more absolute or relative moral code (Baghramian and Carter 2019). Some respondents show greater flexibility in whether an action can be justified, while other show more stringent moral orientations, deeming that an action can never be justified regardless of the circumstances. This series includes questions on whether abortion (V204), violence against other people (V210), avoiding a fare on public transport (V199), and claiming government benefits to which you are not entitled (V198) are ever justifiable. All variables of this series that are included in the overall network appear in this block. V187 asks about the justifiability of action on the part of a nation, whether there are times when it is necessary to go to war.

Table 7. Survey questions: Block 3 – Absolute/Relative Moral Orientation

| Variable | Survey Question | Node Strength |
|---|---|---|

| V204 | [T]ell me…whether you think [abortion] can always be justified, never be justified, or something in between | 9.445 |
|---|---|---|
| V66 | [W]ould you be willing to fight for your country? | 8.734 |
| V100 | In the long run, hard work usually brings a better life – Hard work doesn't generally bring success; it's more a matter of luck and connections | 8.699 |
| V60 | Priority for the country – first choice: A high level of economic growth; Making sure this country has strong defense forces; Seeing that people have more say about how things are done at their jobs and in their communities; Trying to make our cities and countryside more beautiful. | 8.678 |
| V182 | To what degree are you worried about [not being able to give my children a good education] | 8.622 |
| V45 | When jobs are scarce, men should have more right to a job than women | 8.375 |
| V39 | Please mention[if] you would not like to have [immigrants/foreign workers] as neighbors | 7.740 |
| V99 | Competition is good. It stimulates people to work hard and develop new ideas – Competition is harmful. It brings out the worst in people. | 7.506 |
| V148 | Do you believe in God? | 7.201 |
| V187 | Under some conditions, war is necessary to obtain justice | 6.576 |
| V210 | [T]ell me…whether you think [violence against other people] can always be justified, never be justified, or something in between | 6.321 |
| V199 | [T]ell me…whether you think [avoiding a fare on public transport] can always be justified, never be justified, or something in between | 5.972 |
| V41 | Please mention[if] you would not like to have [people of a different religion] as neighbors | 5.645 |
| V198 | [T]ell me…whether you think [claiming government benefits to which you are not entitled ] can always be justified, never be justified, or something in between | 5.471 |
| V128 | Having experts, not government, make decisions according to what they think is best for the country [is a very good way of governing a country] | 3.898 |

In addition to variables of the moral orientation series, the block includes one religious question, namely V148, does the respondent believe in God. The block also includes a question commonly used to measure national pride (V66). Finally, two variables related to competition and success are included in the block: V100, whether hard work usually brings about a better life or whether success is more dependent on luck and connections, and V99, the value of competition.

Block 3 also contains questions that are especially relevant to the dependent variable. The first poses a similar question as the dependent variable, asking whether men should have more right to a job than women when jobs are scarce (V45). The block also contains two questions out of a series related to prejudice and stigma. These variables ask whether the respondent would prefer not to have immigrants and foreign workers as neighbors (V39) as well as whether the respondent would not like to have people of a different religion as neighbors. Other variables in this series are included in the network, but appear in Block 6 and Block 8.

Finally, this block contains the only negative ties between variables within the same cluster. These negative ties appear between V41, whether or not the person would not like to have people of a different religion as a neighbor, and V210 and between V41 and V199. V210, whether violence against other people is justifiable, and V199, whether avoiding a fare is justifiable, are strongly related, with an dot product of 0.867, as seen in Appendix G. V41 and the two variables are negatively correlated, though to a fairly weak degree. The dot product between V41 and V199 is -0.133, while the dot product between V41 and V210 is negligible at -0.002.

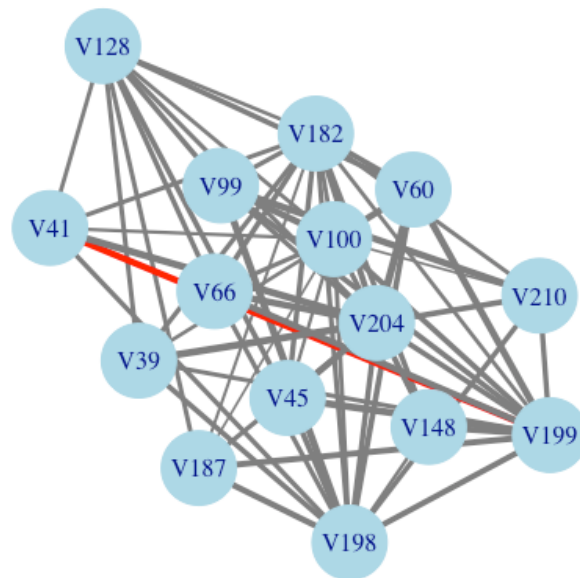

Figure 8. Network of variables in Block 3
Positive ties represented by grey links; negative ties by red ties.

Figure 8 shows the variable network in Block 3. Positive ties between variables are represented by grey edges, while negative ties are represented by red ties. Furthermore, the edge weight is represented by the width of ties, with edges of larger weights having greater thickness. While two negative ties appear in the network, the edge weight of the link between V41 and V210 is so low that the edge is not visible in this network.

The keyword for Block 4 is democracy, taken from the V140, the most central node in Block 4. This variable asks the respondent if it is important to them to live in a democratically governed country.

Table 8. Survey questions: Block 4 – Democracy

| Variable | Survey Question | Node Strength |
|---|---|---|
| V140 | How important is it for you to live in a country that is governed democratically | 3.051 |
| V218 | Indicate whether you use [printed magazines] to obtain information | 2.642 |
| V101 | People can only get rich at the expense of others – Wealth can grow so there's enough for everyone | 2.418 |
| **V55** | [H]ow much freedom of choice and control [do] you feel you have over the way your life turns out? | 2.262 |
| **V97** | Private ownership of business and industry should be increased – Government ownership of business and industry should be increased | 1.345 |
| **V71** | It is important to this person to be rich; to have a lot of money and expensive things | 1.298 |

The other variables in the cluster all show the highest degree of similarity with the question of democracy, implying that views about importance of wealth and how it can be accumulated (V71 and V101), how much freedom of choice and control the respondent feels (V55), and government ownership of businesses (V97) are correlated with the overall importance of democracy to a person. V71, whether it is important for the respondent to be rich, is the only exception, as it shows a very small degree of greater similarity to questions about how wealth can be accumulated. Whether or not democracy is important for respondents shows a high degree of similarity to how respondents obtain information, as represented by V218.

As the Block 4 network shown in Figure 9 demonstrates, the block is a clique. All variables are correlated with one another, to greater or lesser degrees as depicted by the varying edge widths.

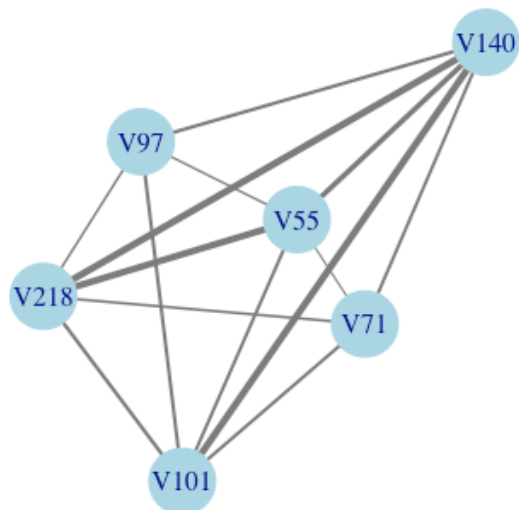

Figure 9. Network of variables in Block 4
Positive ties represented by grey links; negative ties by red ties.

Block 5 comprises two questions. One asks the respondents whether they agree with the statement that science and technology is improving life (V192). The second question asks what is their respondents' priority for the country (V61). This variable is part of a two-part series which asks about priorities for countries. The related variable, V60, asks respondents what their first priority for a country is.

Table 9. Survey questions: Block 5 – Science and Technology

| **Variable** | **Survey Question** | **Node Strength** |
|---|---|---|
| **V192** | Science and technology are making our lives healthier, easier, and more comfortable | 0.260 |
| **V61** | Priorities for the country- second choice: A high level of economic growth; Making sure this country has strong defense forces; Seeing that people have more say about how things are done at their jobs and in their communities; Trying to make our cities and countryside more beautiful. | 0.260 |

Block 6 contains four out of the remaining five questions in the series related to prejudice and stigma. These variables ask whether the respondent would not like to live near drug addicts (V36), people with AIDS (V38), homosexuals (V40), and heavy drinkers (V42). Only one other variable is included in the block, which asks the respondent if they identify as a religious person (V147).

Table 10. Survey questions: Block 6 – Prejudice and Stigma

| **Variable** | **Survey Question** | **Node Strength** |
|---|---|---|
| **V36** | Please mention[if] you would not like to have [drug addicts] as neighbors | 3.246 |
| **V38** | Please mention[if] you would not like to have [people who have AIDS] as neighbors | 3.182 |
| **V40** | Please mention [if] you would not like to have [homosexuals] as neighbors | 2.941 |
| **V42** | Please mention[if] you would not like to have [heavy drinkers] as neighbors | 2.863 |
| **V147** | Independently of whether you attend religious services or not would you say you are [a religious person] | 2.269 |

As Figure 10 show, the block is a clique, and all variables are correlated with one another.

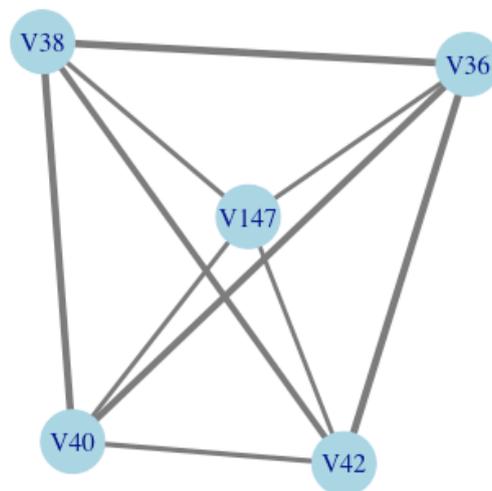

Figure 10. Network of variables in Block 6
Positive ties represented by grey links; negative ties by red ties.

Block 7 represents the only singleton block, meaning the block comprises only one variable. This variable, V150, asks whether the basic meaning of religion is to follow norms or to do good to other people. Here, the survey the respondent to consider the meaning and value of religion for its dogmatic or practical qualities. This variable attempts to understand how religion motivates behavior.

Table 11. Survey questions: Block 7 – Religion (and behavior)

| Variable | Survey Question |
|---|---|
| **V150** | The basic meaning of religion is to follow religious norms and ceremonies [or] to do good to other people |

Block 8 comprises two questions, the first about religion (V151) and the second regarding stigma surrounding unmarried couples (V48). V151 asks whether the basic meaning of religion is to make sense of this life or of life after death, resembling the singleton variable of V150. In this case, the respondent is asked for the role of religion in understanding the universe and human experience, characterized as both life before and after death. Finally, this block includes the final variable on the series of prejudice and stigma, in this case towards unmarried couples living together.

Table 12. Survey questions: Block 8 – Religion 2 (and cognition)

| Variable | Survey Question | Node Strength |
|---|---|---|
| **V151** | The basic meaning of religion is to make sense of life after death [or] to make sense of life in this world | 0.347 |
| **V43** | Please mention[if] you would not like to have [unmarried couples living together] as neighbors | 0.347 |

Having classified each block according to the keyword, it is possible to arrange the determinants into a network graph, showing the positive or negative relationships between each determinant. This network is shown in Figure 11.

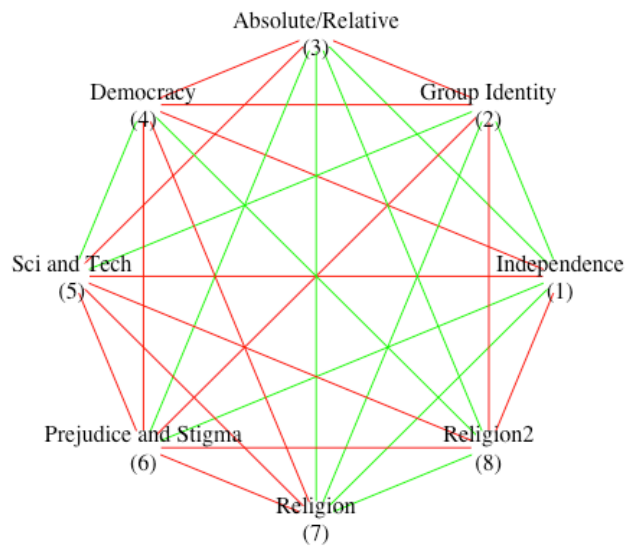

Figure 11. Network of determinants
Green ties represent positive edges; red edges represent negative ties. Edges have no weight.

The edges between determinants can have negative or positive signs, as shown by the red and green edges respectively. The edges have no weights.

# 6 Discussion
## 6.1 Countries

The nine countries selected have greater similarities than differences in their determinants of attitudes towards immigrants, as seen by the network graph and by the positive ties between all countries. The countries form a clique; each country has a positive degree of similarity to all other countries. Dividing the network into communities decreased the overall modularity of the clusters, indicating that all nodes belonged in a single cluster. Thus, there does not appear to be any significant difference in the determinants of attitudes between countries of different development levels, migration profile, or between East Asian and Southeast Asian countries.

While countries can be considered more similar than dissimilar, the slight variations within the group will be discussed. Thailand and Malaysia have the highest node strength of the group. This high node strength is due to the large degree of similarity they have with each other, as well as the strong ties between Malaysia and Hong Kong and between Thailand and Taiwan. The four countries form a strongly linked clique. These countries are have net positive immigration, meaning a greater number of people enter than country than leave.

In contrast to this tightly connected clique, the Philippines shows the lowest degree of similarity with other countries and is the most unique in their determinants of attitudes. This low degree of similarity may reflect the unique circumstances of Philippine migration.

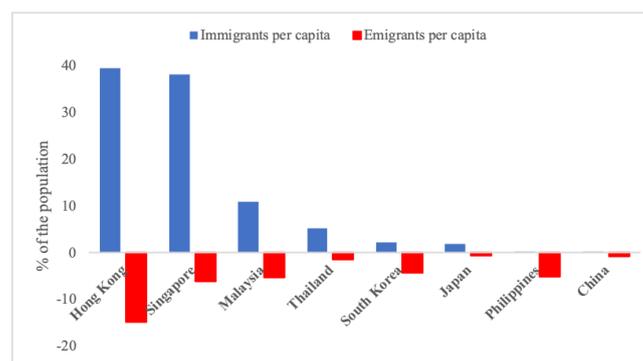

Figure 12. Immigrants and emigrants per capita.
Data source: UN DESA, 2019 and World Bank, 2019.

As shown by Figure 12, immigrant stock per capita is almost non-existent. This level of immigration is very low for the sample. In terms of immigrant stock per capita, the Philippines' low numbers are only comparable to China, a country with a population 10 times greater than the Philippines. Furthermore, as Figure 13 shows, a much larger proportion of immigrants come from Europe and North America, an atypical trend in the sample.

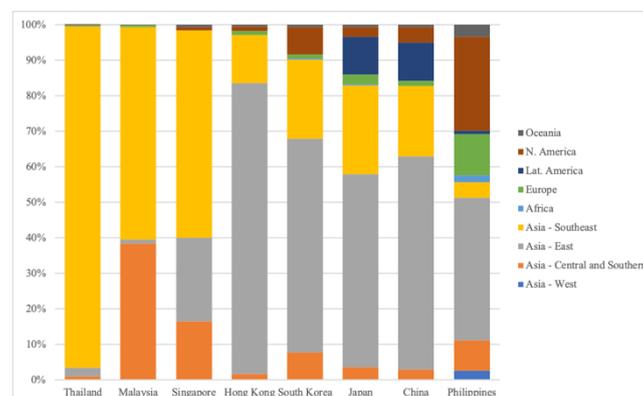

Figure 13. Origins of immigrants, percentage of immigrant stock.

Data source: UN DESA 2019a.

While scholars generally focus on the role of immigration profile in determining attitudes towards immigrants, in the case of the Philippines, its emigration profile seems equally relevant. The emigration of Filipinos, and especially Filipina women, has long been an integral part of the Filipino economy and society, in what Massey calls "a culture of migration" (Massey, 1993, pp. 452). In terms of absolute numbers, the Philippines sends the most emigrants abroad in the sample, after China (UN DESA 2019a). Emigrants from the Philippines also tend to go to locations that are atypical as compared to other countries in the sample, such as Gulf States, as Figure 14 shows. The unique history of migration in the Philippines, its small and atypical immigrant stock, as well as its large emigrant population in destinations that are uncommon for this sample, likely contribute to its low similarity with other countries in this sample.

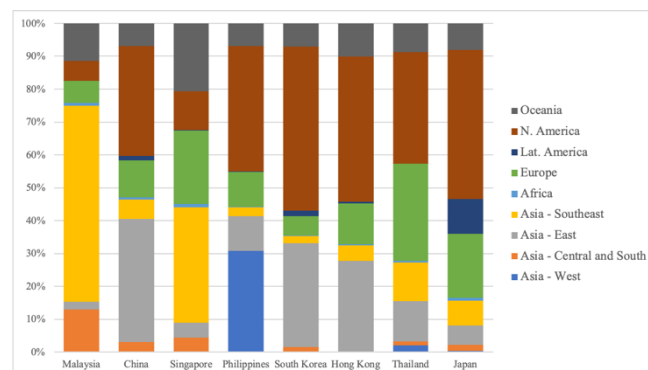

Figure 14. Destinations for emigrants, percentage of immigrant stock.
Data source: UN DESA 2019a.

Japan also reaffirms the importance of emigration profile in determining attitudes towards immigrants. Japan represents a unique case, as the country with the second smallest node strength. It most closely resembles Korea in its determinants of attitudes. Both are East Asian, high-income countries with similar populations of immigrants per capita. However, in its determinants of attitudes, Korea more closely resembles countries like Malaysia and Taiwan. Why this difference exists may lie in the differences in the migration profile of the two countries. First, compared to the majority of countries in the sample, Japan receives a greater proportion of immigrants from Latin America and the Caribbean. Second, the per capita emigrant population in Japan compared to Korea is several times smaller as compared to Korea. Furthermore, proportional to other destinations, very few emigrants from Japan choose to move to East Asia countries as compared to Korean emigrants. Rather, Japanese emigrants tend to move to Western countries and Southeast Asian countries. The relative similarity of the countries in their immigration profile, but the large differences in their emigration profile suggest that a country's history of emigration may be equally relevant area of study for the determinants of attitudes towards immigrants.

From this study, it can be concluded that countries in East and Southeast Asia show a significant degree of similarity in their determinants towards immigrants. No differentiation was found between countries in Southeast Asia and East Asia. When comparing the countries within the cluster, it is clear that a group of the net receivers of migration, Thailand, Malaysia, Hong Kong, and Taiwan, create a cohesive clique with a large proportion of heavily weighted edges. On the other hand, two countries with unique emigration profiles, the Philippines and Japan, show the least similarity to countries in the sample. These results

Network Analysis of Attitudes towards Immigrants in Asiaimply that a country's emigration profile is significant in determining attitudes towards immigrants at home.

## 6.2 Variables

The findings of this study have implications for the prevailing theories of how attitudes towards immigrants form and the applicability of these theories to countries outside European and Settler countries. Two determinants, science and technology and absolute/relative moral orientation, had not been mentioned by the literature review and warrant further research. Other determinants had been mentioned by the previous literature but show different relationships than had been hypothesized. The results of this study support, contradict, and expand theories related to determinants of attitudes, specifically the effects of education, social class, group identity, religion, prejudice, and anxiety on attitudes towards immigrants.

Despite its strong support in the literature, education was not found to be a significant variable in determining attitudes towards immigrants for any countries in the sample. As such, the variable measuring the highest education attained by respondents was not included in the variable network. This finding echoes the results of Meseguer and Kemmerling's (2016) study, which found a limited effect of education on attitudes in Latin American countries. These findings suggest that education's effects are strongest in European and Settler countries and become less applicable in other regions of the world. Exactly why education loses its predictive power outside of the West requires further investigation into the different ways in which higher education functions in the economic systems of different countries as well as how the education system itself instills or correlates with other cultural values related to attitudes towards immigrants.

Unlike education, subjective social class was included in the variable network. For most countries in the sample, social class had no significant relationship with the dependent variable and, as such, its exact relationship with attitudes towards immigrants cannot be concluded. For the countries in which social class did have a significant effect, the sign of the coefficient was not consistent. In China and Singapore, respondents who identified as belonging to lower social classes were more likely to express negative attitudes towards immigrants. This result would appear to support labor market competition theories in which people of lower classes and who are in competition with immigrants for jobs would have more negative attitudes towards immigrants. However, in Thailand, respondents in higher classes were more likely to express negative attitudes towards immigrants. Social class encompasses not only income but also non-material elements of prestige and social standing. Its inclusion in Block 1 (independence and social dependencies), the high degree of similarity to variables related to institutional trust, as well as the absence of income in the network of variables implies that the effect of social class on attitudes is more closely related to the social standing aspects of class than its economic dimensions. Moreover, the differing signs between countries in which class had a significant effect implies that class may have a mediating effect on attitudes, but that it depends on the overall social and economic context of the country. This study can, thus, conclude that the effect of social class appears to be more closely related to social standing rather than the material aspects of class and that the effect of class on attitudes depends on the social and economic context of a country.

Theories of prejudice often cite social identity theory to explain how group identities affect individuals' prejudice against immigrants and minorities. Both group identities and prejudice were identified as determinants and were negatively correlated, though weakly. This finding suggests that greater attachment to one's local community or to one's regional organization was correlated with greater prejudice. However, for most countries, the relationship between group identity and negative attitudes towards immigrants was not statistically significant. Thus, while it does appear that prejudice and group identity have a negative relationship, this relationship requires further investigation due to the small sample size.

For the countries for which there was a significant relationship between group identity and negative attitudes towards immigrants, the signs of these relationships were not consistent. Respondents in Singapore who saw themselves as more closely tied with either their local community or the regional association were more likely to have positive attitudes towards immigrants, while respondents in the Philippines who felt more closely tied to both groups were more likely to have negative attitudes towards immigrants. Both forms of attachment caused respondents in the Philippines to become more negative towards immigrants, while the opposite trend emerged in Singapore. As both forms of attachment behave very similarly, it appears that the object of attachment is not nearly as important as the strength of the attachment. However, due to the inconsistency of the sign, how group identities affect attitudes towards immigrants appears to be country-specific. As these variables were significant for few countries in the sample, the exact relationship between attitudes and group identities cannot be concluded from this study.

Religion plays a diffuse role in determining attitudes towards immigrants, with variables measuring different aspects of religion having very different relationships with other variables. All four variables related to religion fell into different blocks. Overall, questions regarding religion show a negative, though very weak, correlation with stronger group identity (Block 2), democracy (Block 4), and science and technology (Block 5). In their relationships with other clusters, the major distinction between variables related to religion lies in their interaction with Block 1 (independence and social dependencies) and Block 6 (prejudice and stigma). In other words, what facet of religion is being discussed becomes salient only with regards to prejudice and stigma, and independence. Belief in God as well as whether the respondent identifies as religious does impact people's prejudices and stigmas; however, how one interprets the meaning of religion has no bearing on prejudices and stigmas. Furthermore, whether or not one believes in God does not affect whether one believes that the meaning of religion is to follow norms or to do good towards others, but is correlated with all other variables regarding religion.

Stigma and prejudice against foreign workers and people of a different religion appear to be qualitatively different from prejudice against other groups. As the previous paragraph mentions, the major causes of differentiation between variables in this series are their relationships with Block 7 and Block 8, determinants related the meaning of religion. Whereas variables in Block 6 have negative edges with Block 7 and Block 8, V41, which asks whether the respondent would like to have neighbors of a different religion, shows a fair amount of similarity with both; the inner products of the relationship of V41 with V150 and V151 are 0.346 and 0.228, respectively. V39, which asks about prejudice and stigma against immigrants, has a lesser degree of dissimilarity from Block 6, as it has a weakly negative inner product with V150 and a weakly positive inner product with V151. This pattern suggests that prejudice against people of different religions or against immigrants correlates with the person's conception of the meaning of religion. However, stigma and prejudice towards other groups, such as drug addicts, homosexuals, or people of another religion, is not correlated with the meaning of religion.

Finally, this study finds that variables related to anxiety are located in blocks related to independence and absolute/relative moral orientations, Blocks 1 and 3 respectively. In their study on the role of anxiety in information seeking about immigration, Gadarian and Albertson identified four major causes of concern: economic concerns, concerns about exploiting the social welfare system, cultural worries, and security concerns (2014). The placement of variables related to anxiety in certain blocks suggest that anxiety may operate on certain considerations to a larger extent than others. In particular, Block 3 relates to the absolute and relative moral orientation of respondents. The placement of the fear here may suggest that affronts to permissible behavior creates greater feelings of threat than other considerations. This finding helps explain the precedents of the much more stringent and punitive reactions to illegal immigration than documented immigration (Hood and Morris 1998). Variables measuring anxiety in Block 1, which measures independence as well as institutional trust and normative behavior, reinforce this interpretation. These findings suggest that fears about the cultural threat posed by immigrants' inclusion into society may have less to do with group identity and prejudice than a more rigid adherence to established, normative behavior. Interestingly, anxiety was not included in clusters regarding prejudice, but these variables measuring anxiety did show a positive correlation with the series of variables regarding prejudice and stigma in both Blocks 3 and 6. Furthermore, anxiety has a somewhat mixed relationship with group identity, with concerns about providing a good education for their children showing negative edges and other measures of anxiety showing a very low degree of similarity. These findings suggests that anxiety does have a galvanizing effect on prejudices, but that it is moderated through anxieties related to permissible behavior.

## 7 Conclusion

This study has found that overall, the countries in the sample represent a cohesive group, showing greater similarities than differences. However, given the countries which are the most unique within the sample, the Philippines and Japan, emigration profile emerges as an area of study that could be pertinent to how attitudes towards immigrants form and which has thus far been largely neglected by the literature review.

The one-mode projection of the variables produced eight determinants of attitudes: independence, group identity, absolute/relative moral orientation, democracy, science and technology, prejudice and stigma, and two determinants related to the meaning of religion. Some of these determinants had not been previously identified by research, including science and technology and absolute/relative moral orientation.

The relationships between variables, as well as what variables were included in each cluster, have complicated and expanded the previous research on determinants. Several variables which were assumed to be relatively consistent in their predictive power were found to be insignificant, as in the case of education, or inconsistent in the direction of the effects, as in the case of class. In particular, the absence of education as a significant variable stresses the need for greater research into countries outside of Europe and the Settler countries. This finding complicates the proposed relationship between education and labor market competition and reinforces the hypothesis that higher education in European and Settler countries correlates with cultural values that liberalize attitudes towards immigrants. However, before concluding that education's effect on attitudes occurs because the curriculum is inspiring greater critical thinking in its students or through self-selection, it is necessary to evaluate how education mediates other socio-economic aspects of life. The meaning of education, as a class signifier, a determinant of labor market outcomes, and its relationship with, depends on the country context and could conceivably be related to macro-level factors related to inequality or social mobility. Despite the abundance of studies which include education, the exact relationship between education and attitudes towards immigrants requires further research. Equally, the effect of class appears to be country-specific, as seen by the differing signs of its coefficients. Future research would do well to investigate under what conditions members of lower classes feel antagonism towards immigrants, as in the case of China and Singapore, rather than solidarity with immigrants, as in the case of Thailand.

Supporting social identity theory, stronger group identities were found to be correlated with prejudice; though due to the small sample size, this trend cannot be confirmed.

Interestingly, the object of attachment, whether it was one's local community or a regional organization, appeared to be less significant than the strength of the attachment, a finding that should be confirmed and expanded upon in future research.

In this study, prejudice and religion appeared to be interrelated, as variables in these series were differentiated due to their relationship with variables related to religion. In particular, variables related to prejudice against immigrants/foreign workers and people of different religions were separated from variables related to prejudice against groups, like drug users, people suffering from AIDS, and so on, because of their relationships with variables related to the meaning of religion. Thus, mentioning and discussing religion is not necessarily in conflict with discussion regarding prejudice; in fact, the two can be mutually reinforcing when speakers carefully choose which facet of religion to discuss. Stressing religious identification and belief in God may activate prejudice, while discussions of the meaning of religion may mitigate it.

Finally, anxiety, which had been identified by previous research as a salient determinants, is found to operate specifically on people's expectations of normative social behavior, as seen by its inclusion in Block 3 (absolute/moral orientation) and Block 1 (independence and social dependencies). As anxiety appears to be correlated with expectations of how others should behave and whether one has a greater or lesser ability to accept non-normative behavior, media likely plays a large role in moderating or exacerbating fears. Policymakers and others may find depictions of the mundane social expectations and rules of immigrant and autochthonous populations to be fruitful pathways to quell fears related to immigration.

While the relationships of these determinants and the macro-level factors that may be underlying them cannot be concluded in all cases, this study hopes to inspire greater research into these understudied areas and to expand the field. Research should reflect the changing nature of migration, including expanding into understudied regions, and should employ new techniques, such as network science, where applicable. While methodological difficulties exist, identifying generalizable variables and the macro-level economic and social conditions that mediate them can only enrich understanding of when and why a society accepts or rejects newcomers.

**Abbreviations**
ATII: Attitudes towards immigrants and immigration; CAN: Causal Attitude Network; ESS: European Social Survey; ISSP: International Social Survey Programme; LAPOP: Latin American Public Opinion Project; UN DESA: United Nations, Department of Economic and Social Affairs; WVS: World Values Survey

**Availability of data and materials**
The datasets analyzed during the current study are available in the World Values Survey repository (Ingelhart et al. 2014), the UN DESA website (UN DESA 2019a), and the World Bank website (World Bank 2019).

**Competing interests**
None.

**Funding**
The present study was supported by the Ministry of Education, Culture, Sports, Science and Technology (MEXT), Grants-in-Aid for Scientific Research (B), Grant No. 17KT0034 (2017-2019).

**Authors' contributions**
RKK provided key contributions in performing the survey of the related work, the regression analysis, and network analysis. YI supervised the project and conceptualized the bipartite graph expression of the regression analysis. All authors read and approved the final manuscript.

**Acknowledgements**
We are grateful to D. Ialnazov and M. Deroche for their insightful comments and encouragement.